\documentclass[pre, aps,groupaddress, showpacs, twocolumn]{revtex4-1}
\usepackage{epsfig,amsmath,graphicx,amssymb,overpic}
\def\be{\begin{equation}}
\def\ee{\end{equation}}
\def\bee{\begin{eqnarray}}
\def\ene{\end{eqnarray}}
\def\bes{\begin{subequations}}
\def\ees{\end{subequations}}

\begin{document}
\title{Matter-wave solutions in the Bose-Einstein condensates with \\ the harmonic and Gaussian potentials}
\author{Zhenya Yan$^{1}$}
\email{zyyan@mmrc.iss.ac.cn}
\author{Dongmei Jiang$^{1,2}$}

\affiliation{$^1$Key Laboratory of Mathematics Mechanization,
Institute of Systems Science, AMSS, Chinese Academy of Sciences,
Beijing 100190, China\\
$^2$Department of  Mathematics, Qingdao University
of Technology,  Qingdao 266033, China\vspace{0.1in}}

\date{\vspace{0.1in} November 2011, Phys. Rev. E   {\bf 85}, 056608 (2012)}


\begin{abstract}
We study exact solutions of the quasi-one-dimensional Gross-Pitaevskii (GP) equation with the (space, time)-modulated potential and nonlinearity and the time-dependent gain or loss term in Bose-Einstein condensates. In particular, based on the similarity transformation, we report several families of exact solutions
of the GP equation in the combination of the harmonic and  Gaussian potentials, in which some physically relevant solutions are described. The stability of the obtained matter-wave solutions is addressed numerically such that some
stable solutions are found. Moreover, we also analyze the parameter regimes for the stable solutions. These results may raise the possibility of relative experiments and potential applications.

\end{abstract}


\maketitle

\vspace{0.2in}

\baselineskip=12pt

\section{Introduction}

The nonlinear Schr\"odinger (NLS) equation (alias the Gross-Pitaevskii (GP) equation in Bose-Einstein condensates)
 is one of the most useful physical models,  which actually arises in many physics fields, such as nonlinear quantum field theory, condensed matter, plasma physics, nonlinear optics, photonics, fluid mechanics, semiconductor
electronics, phase transitions, biophysics, star formation, econophysics, and so on (see, e.g., \cite{b1,b2,b3,b4,b5,b6}). The first exact solutions of the one-dimensional (1D) NLS equation were obtained by using the inverse scattering method~\cite{bb01}, and over these years there have been many significant contributions to the studies of exact solutions for some extensions of the NLS
 equation~\cite{bb02,bb03,bb04,j8}.

 Recently, exact localized solutions of the GP equations with  varying coefficients  have drawn much attention due to their potential applications in the theory of Bose-Einstein condensates (BECs)~\cite{j5,j55,j6,j7,fr}. It is in general difficult to analytically solve these GP equations due to their nonlinear nature, but there have been some works investigating exact solutions of the GP equations with (time, space)-modulated potential and nonlinearity, such as the harmonic potential
$x^2$~\cite{liang, bb07,bb08, yan11}, the van der Waals law potential $1/x^3$~\cite{yu9}, the harmonic potentials in the three-dimensional space~\cite{yan06}, in the $d$-dimensional space~\cite{vvk}, as well as in the bounded two-dimensional domains~\cite{yan12}.

For the usual 3D external potential $V_{\rm ext}(x,y,z)=\frac{1}{2}m(\omega_xx^2+\omega_yy^2+\omega_zz^2)$ in BECs, where $m$ is the atomic mass and  the trap frequencies $\omega_{x, y, z}$ along the three directions are in general different and
may be used to control the shape of the condensates. The isotropic case $\omega_x=\omega_y=\omega_{\bot}\approx\omega_z$ denotes that the 3D BEC is almost spherical, but the other two weakly anisotropic cases $\{\omega_{\bot}<\omega_z \,\, {\rm or} \,\, \omega_{\bot}>\omega_z\}$ imply that the 3D BEC is cigar shaped or disc-shaped, respectively. Moreover the strongly anisotropic cases $\{\omega_{\bot}\gg\omega_z \,\, {\rm or} \,\, \omega_{\bot}\ll\omega_z\}$ are related to the lower-dimensional BECs, i.e., the effectively quasi-1D or
 quasi-2D BECs, respectively~\cite{b5}. The lower-dimensional BECs in highly asymmetric traps have been verified in the study of the Bosonic and Fermionic gases~\cite{low}.

 Here we consider a cigar-shaped BEC of a relatively low density, when the energy of two body interactions is much less than the kinetic energy in the transverse direction, i.e., $|a_s|\mathcal{N}\ll a_\bot$ with $\mathcal{N}$ being a total number of atoms and $a_\perp=\sqrt{\hbar/(m\omega_\bot)}$~\cite{j55}, the macroscopic wave function of a quasi-one-dimensional condensate in BECs obeys the effective 1D GP equation~\cite{j5,b5,j55,j6}
\bee \label{nls0}
 i\hbar\frac{\partial\!\Psi}{\partial t}
  \!\!=\!\!\left[\!-\frac{\hbar^2}{2m}\frac{\partial^2}{\partial x^2}\!+\!\mathcal{V}_{\rm ext}(x,\!t)
   \!\!+\!\mathcal{G}_{\rm 1D}(x,\!t)|\Psi|^2\!\!+\!i\vartheta(t)\!\right]\!\!\Psi,\,\,\,\,
 \ene
where $\Psi\equiv \Psi(x,t)$ denotes the condensate wave function, $m$ is the atomic mass, $\mathcal{V}_{\rm ext}(x,\!t)$ denotes the time- and space-modulated external confining potential, which can be usually chosen in the form of the harmonic well or the optical lattice~\cite{j5}, the 1D effective coupling coefficient
$\mathcal{G}_{\rm 1D}(x,t)=2\hbar\omega_{\bot}a_s(x,t)$ stands for a measure of the nonlinear two-body interactions between the atoms in quasi-1D geometries with strong transverse confinement with $\omega_{\bot}$ being the transverse trapping frequency and $a_s(x,t)$ being the time-dependent
$s$-wave scattering length for elastic atom-atom collisions modulated by a Feshbach resonance~\cite{j5,fr}, in which the scattering length
$a_s(x,t)$ is a function of the varying magnetic field $B(x,t)$:
 \bee\nonumber
  a_s(x,t)=\hat{a}\left(1+\frac{\Delta}{B_0-B(x,t)}\right),
 \ene
where $\hat{a}$ is the value of the scattering length far from resonance, $\Delta$ represents the width of
the resonance, $B_0$ is the resonant value of the magnetic field, and $B(x,t)$ is external (space, time)-modulated magnetic field~\cite{fr}. The scattering length
$a_s$ can be chosen as either positive or negative values corresponding to the attractive interactions (e.g., for
$^{7}${\rm Li} or $^{85}${\rm Rb} in the BECs) or repulsive interactions (e.g., for $^{87}${\rm Rb} or $^{23}${\rm Na} in the BECs) between the atoms~\cite{b5,j5, j55,j6}.  The time-dependent gain or loss term $\vartheta(t)$ corresponds to the mechanism of continuously loading external atoms into the BEC (gain case) by optical pumping or continuous depletion of atoms from the BEC  (loss case)~\cite{gl}.

Normalizing the density $|\Psi|^2$, length, time and energy in Eq.~(\ref{nls0}) in units of  $2a_s,\, a_{\bot}=\sqrt{\hbar/(m\omega_{\bot})},\, \omega_{\bot}^{-1}$, and $\hbar\omega_{\bot}$, respectively, we obtain a 1D effective GP equation with the
space- and time-modulated potential and nonlinearity and the time-dependent gain or loss term in the dimensionless form~\cite{b5,j5,j55,j6,bb08}
\bee \label{nls1}
 i\frac{\partial\psi}{\partial t}
 =\left[-\frac{1}{2}\frac{\partial^2}{\partial x^2}+V(x,t)+g(x,t)|\psi|^2+i\Gamma(t)\right]\!\psi,\,\,
 \ene
where the external potential $V(x,t)$, the nonlinearity $g(x,t)$, and the gain or loss distribution $\Gamma(t)$ are related to the functions $\mathcal{V}_{\rm ext}(x,t)$,\, $\mathcal{G}_{\rm 1D}(x,t)$, and $\vartheta(t)$ in Eq.~(\ref{nls0}), respectively.  Eq.~(\ref{nls1}) is
associated with the Euler-Lagrange equation $\delta \mathcal{L}/\delta \psi^\ast=0$ in which the Lagrangian density can be written as~\cite{lan,yan06}
\bee
\mathcal{L}=i\left(\psi \frac{\partial\psi^\ast}{\partial t}
  -\psi^\ast\frac{\partial\psi}{\partial t}\right)\!+\left|\frac{\partial\psi}{\partial x}\right|^2 \nonumber
  \qquad\quad\quad\, \\
  +g(x,t)|\psi|^4+2[V(x,t)+i\Gamma(t)]|\psi|^2,
\ene
where $\psi^\ast\equiv \psi^\ast(x,t)$ denotes the complex conjugate of the complex field $\psi(x,t)$. Eq.~(\ref{nls1}) with the harmonic potential $V(x,t)=a(t) x^2$ has
been shown to possess exact solutions only for the attractive
interaction $g(x,t)<0$~\cite{bb08}.

In this paper, to translate our results into units relevant to the experiments~\cite{exp00,exp0, exp1,exp2}, we perform this protocol for the attractive case by preparing BECs of up to $\sim 10^4-10^5$ $^{7}${\rm Li} atoms with $m=11.65\times 10^{-27} {\rm Kg}$ and $\omega_{\bot}=2\pi\times 10^3${\rm Hz}, in which a unity of the dimensionless space corresponds to $a_{\bot}=\sqrt{\hbar/(m\omega_{\bot})}\simeq 1.19{\rm \mu m}$
and a unity of the dimensionless time corresponds to $1.6\times 10^4\,{\rm s}$. For the repulsive interaction, we prepare BECs of up to $\sim 10^5$ $^{87}${\rm Rb} atoms with $m=1.44\times 10^{-25} {\rm Kg}$ and a unity of the dimensionless space corresponds to $a_{\bot}=\sqrt{\hbar/(m\omega_{\bot})}\simeq 0.34{\rm \mu m}$. Here, we want
to point out that in the following discussion, both the space and time extents of the obtained solutions are both in the range of quasi-one-dimensional validity~\cite{j5,b5,j55,j6}. We consider spatially localized solutions of
Eq.~(\ref{nls1}) with two kinds of interactions (the repulsive interaction $g(x,t)>0$ and the attractive
interaction $g(x,t)<0$) and the combination of the harmonic and Gaussian potentials $V(x,t)$. To do this, we connect exact solutions of the GP equation with space- and time-modulated potential and nonlinearity given by Eq.~(\ref{nls1}) with those of the stationary GP equation (\ref{snls1}) by using the proper similarity transformation and some constraints about the varying coefficients. In addition, the stability of the solutions and the effects of amplitude and phase noises for the solutions are analyzed via the numerical simulations.

The paper is organized as follows. In Section II, we describe the similarity transformation and choose proper variables in order to obtain the combination of the harmonic and Gaussian potentials. In Sections III and IV, we present several families of exact solutions of the GP equation with both the stationary and the (space, time)-modulated potentials and nonlinearities. Moreover, we also study the stability of the obtained solutions numerically to find some stable solutions. In particular, we analyze the effects of both the amplitude noise and the phase noise for the stable solutions. We also give the parameter regimes for the stable solutions. These results may raise the possibility of relative experiments and potential applications in BECs and other related fields. Finally, the results are summarized in Section V.

\section{Similarity solutions}

\subsection{Similarity transformation and constraints}

To seek spatially localized solutions of Eq.~(\ref{nls1}) satisfying the condition
$\lim_{|x|\rightarrow \infty}|\psi(x,t)|=0$, we choose the complex field $\psi(x,t)$ in the form
\bee
 \psi(x,t)=\rho(x,t)e^{i\varphi(x,t)}\Phi[X(x,t)]  \label{trans}
 \ene
 and require that the real-valued function $\Phi[X(x,t)]$ solves the 1D stationary GP equation with constant coefficients
\bee
\varepsilon \Phi(X)=-\frac{d^2\Phi(X)}{dX^2}+G\,|\Phi(X)|^2\Phi(X), \label{snls1}
\ene
whose solutions can be obtained by using some transformations (see, e.g., \cite{yan06, yan}). Here $X\equiv X(x,t)$ denotes the similarity variable and is a real function of space $x$ and time $t$ to be determined below, $\varepsilon$ is the eigenvalue of the nonlinear stationary equation, and $G$ is a constant interaction ($G>0$ corresponds to the repulsive interaction and $G<0$ corresponds to the attractive interaction). The amplitude $\rho$ and phase $\varphi$ are both real functions of $x$ and $t$ to be determined later.

 As a consequence, we find the amplitude $\rho(x,t)$, the similarity variable $X(x,t)$, the phase $\varphi(x,t)$, the external potential $V(x,t)$, the nonlinearity $g(x,t)$, and the gain or loss term $\Gamma(t)$ are required to satisfy the following system of nonlinear partial differential equations
\bes \label{sys1} \bee
\label{sys11} && (\rho^2)_t+(\rho^2\varphi_x)_x-2\Gamma(t)\rho^2=0,\quad (\rho^2 X_x)_x=0,\,\, \\
\label{sys12} && g(x,t)=\displaystyle \frac{GX_x^2}{2\rho^2},\quad X_t+\varphi_xX_x=0,\\
\label{sys13} && V(x,t)=\frac{1}{2}\left(\frac{\rho_{xx}}{\rho}-\varphi_x^2-\varepsilon X_x^2\right)-\varphi_t.
\ene\ees

Since the solutions of the reduced equation (\ref{snls1}) are known, thus if we can determine these variables $X(x,t),\, \rho(x,t),\, \varphi(x,t),\, g(x,t)$, and $V(x,t)$ from system (\ref{sys1}), then we can obtain the solutions of Eq.~(\ref{nls1}) by using those of the reduced equation (\ref{snls1}) and the similarity transformation (\ref{trans}).

It may be difficult to directly solve system (\ref{sys1}). Here we only need its some nontrivial solutions. To solve system (\ref{sys1}), we choose the similarity variable $X(x,t)$ in the form  $X(x,t)=F(\xi)$, where $\xi(x,t)=\gamma(t)x+\delta(t)$ and $F(\cdot)$ is a differentiable function.  Here $\gamma(t)$ is
the inverse of the width of the localized solution, and $-\delta(t)/\gamma(t)$ is the position of its center of mass. After some algebra it follows from system (\ref{sys1}) that
 \bes \bee
&&  \rho(x,t)=e^{\int \Gamma(t)dt}\sqrt{\frac{\gamma}{F'(\xi)}}, \\
&&  \varphi(x,t)=-\frac{\gamma_t}{2\gamma}x^2-\frac{\delta_t}{\gamma}x+\alpha(t),\\
&&  g(x,t)=\frac{\gamma G}{e^{2\int \Gamma(t)dt}}F'^3(\xi), \\
  \label{p2}
&&   V(x,t)=\frac{1}{2}\left(\frac{\rho_{xx}}{\rho}-\varphi_x^2-\varepsilon \gamma^2F'^2(\xi)\right)-\varphi_t,
\ene\ees
where the prime denotes the derivative with respect to the variable $\xi$, $\alpha(t)$ is an arbitrary function of time, and the condition $\gamma/F'(\xi)>0$ is required.

\subsection{The choice of the parameters}

Since the choice of the possible function $F(\xi)$ is very rich in the amplitude $\rho(x,t)$, the nonlinearity $g(x,t)$, and the external potential $V(x,t)$, considering the external potential $V(x,t)$ to be a
combination of the harmonic and Gaussian traps and the possible application of our results to BECs~\cite{j5,exp0,gau}, we thus choose the
Gaussian shaped interaction
\bee
 g(x,t)=\frac{\gamma G}{2e^{2\int \Gamma(t)dt}}e^{-3\xi^2}
 \label{nonlinearity}
  \ene
with $\xi=\gamma(t)x+\delta(t)$, which corresponds to $F'(\xi)=\exp(-\xi^2)$, i.e., $F(\xi)=\int^{\xi}_{-\infty}\exp(-s^2)ds$.
It follows from Eqs. (\ref{p2}) and (\ref{nonlinearity}) that we have the external potential
 \bee
 V(x,t)=\frac{p(t)}{2}x^2+q(t)x+h(t)-\frac{\varepsilon}{2}\,\gamma^2(t)e^{-2\xi^2},
 \ene
 where we have introduced the following functions
 \bee \begin{array}{l}
 p(t)=\gamma^4 e^{8\int \Gamma(t)dt}+(\gamma\gamma_{tt}-2\gamma_t^2)/\gamma^2, \vspace{0.1in}\cr
 q(t)=\gamma^3\delta e^{6\int \Gamma(t)dt}+(\gamma \delta_{tt}-2\gamma_t\delta_t)/\gamma^2,\vspace{0.1in}\cr
 h(t)=\gamma^2 e^{4\int \Gamma(t)dt}(1+\delta^2)/2-\delta_t^2/2\gamma^2-\alpha_t.
  \end{array} \ene

 Taking $\delta(t)=0$ and $\alpha(t)=(1/2)\int \gamma(t)^2dt$, then we obtain the required potential, i.e., the combination of the harmonic and Gaussian potential
 \bee
 \label{v} V(x,t)=\frac{p(t)}{2}x^2-\frac{\varepsilon}{2}\,\gamma^2(t)e^{-2\gamma^2(t)x^2}
 \ene
 and other variables
  \bes \label{sys} \bee
  \label{x} &&  X(x,t)=\int^{\gamma(t)x}_{-\infty}e^{-s^2}ds,\\
\label{rho}  && \rho(x,t)=\sqrt{\gamma(t)}\exp\left[\gamma^2(t)x^2/2+\int \Gamma(t)dt\right],\\
  && \label{g} g(x,t)=\frac{G}{2}\,\gamma(t)\exp\left[-3\gamma^2(t)x^2-2\int \Gamma(t)dt\right],\quad \\
  && \label{q1} \varphi(x,t)=-\frac{\gamma_t}{2\gamma}x^2+\frac{1}{2}\int \gamma^2(t)dt.
  \ene\ees

  Thus, if we choose the inverse of the width of the localized solution $\gamma(t)$ and the gain or loss term $\Gamma(t)$, then we can determine these variables $X(x,t)$, $\rho(x,t) $, and $\varphi(x,t)$ for which the solutions of Eq.~(\ref{nls1}) can be obtained from those of Eq.~(\ref{snls1}) using the similarity transformation (\ref{trans}).

  \section{Stationary potential, nonlinearity and solutions}

   In this section we consider the simple case of the stationary nonlinearity $g(x)$ and potential $V(x)$ in Eq.~(\ref{nls1}), which means that $\gamma(t)\equiv \gamma={\rm const}$ and $\Gamma(t)\equiv 0$ \, (i.e., zero-gain or zero-loss term) based on Eq.~(\ref{nonlinearity}), then Eq.~(\ref{nls1}) can be reduced to the GP equation with the external potential and nonlinearity depending only on the spatial coordinate~\cite{bb07}
     \bee \label{nls}
      i\frac{\partial\psi}{\partial t}=\left[-\frac{1}{2}\frac{\partial^2}{\partial x^2}+V(x)+g(x)|\psi|^2\right]\psi.
     \ene
 Thus, based on the results in Section II, Eq.~(\ref{nls}) can be reduced to Eq.~(\ref{snls1}) with $X(x,t)\to X(x)$ via the following similarity transformation (c.f. Eq.~(\ref{trans}))
 \bee
 \psi_{sta}(x,t)=\rho(x)e^{i\varphi(t)}\Phi[X(x)]  \label{trans2}
 \ene
 with the similarity variables being of the form
   \bes \bee
   && \label{xx} X(x)\!=\!\int^{\!\!\sqrt{\omega}x}_{-\infty}\!\!e^{-s^2}\!\!ds
     \equiv \frac{\sqrt{\pi}}{2}\left[1+{\rm erf}(\sqrt{\omega} x)\right],\\
   && \label{rho}\rho(x)\!=\!\omega^{\frac{1}{4}}e^{\frac{\omega x^2}{2}},\,\,\,\, \varphi(t)\!=\!\frac{\omega t}{2} \ene\ees
  with ${\rm erf}(\cdot)$ being the error function, and the external potential and nonlinearity in Eq.~(\ref{nls})  satisfying the conditions
    \bes\bee
    \label{gg} && g(x)=\frac{G\sqrt{\omega}}{2}\,e^{-3\omega x^2},\\
    \label{vv}&& V(x)=\frac{\omega^2}{2}x^2-\frac{\varepsilon \omega}{2}\,e^{-2\omega x^2},
      \ene\ees
  from which the external potential $V(x)$ is a combination of both harmonic potential $(1/2)\omega^2x^2$ and the
  Gaussian potential $-(1/2)\varepsilon \omega e^{-2\omega x^2}$, where $\omega=\gamma^2$. The parameters $\omega^2/2$ and $|\varepsilon \omega/2|$ measure the relative strengths of the harmonic and Gaussian potentials, respectively.

   Notice that when $\varepsilon=0$, the external potential $V(x)$ given by Eq.~(\ref{vv}) becomes the usual harmonic potential and the reduced equation (\ref{snls1}) is changed into the ordinary differential equation
   \bee
   -\frac{d^2\Phi(X)}{dX^2}+G|\Phi(X)|^2\Phi(X)=0,
   \ene
   whose solution has been considered for only the attractive interaction $G<0$~\cite{bb07}. In addition,
   the dynamical evolution of a Fermi super-fluid had been numerically studied based on the quasi-1D density-functional GP equation with a double-well potential similar to Eq.~(\ref{vv})~\cite{pu}.

   In the following, we consider exact solutions and their dynamics of Eq.~(\ref{nls}) with the combination of the harmonic and Gaussian potential (\ref{vv}) and Gaussian shaped interaction (\ref{gg}) for the general case $\varepsilon\omega\not=0$.

\subsection{The repulsive interaction}

We have the periodic sn-wave solution of Eq.~(\ref{snls1}) with the
repulsive interaction $G>0$ ($X(x,t)\to X(x)$)~\cite{yan06}
  \bee \label{SN}
   \Phi_{sn}[X(x)]=\frac{\sqrt{2}\,\nu k}{\sqrt{G}}\,{\rm sn}[\nu X(x),k],
  \ene
  where $k\in (0,1)$, $\nu$ is a constant, $\varepsilon=\nu^2(1+k^2)>0$, $X(x)$ is given by Eq.~(\ref{xx}), the function `${\rm sn}$' denotes the Jacobi elliptic function with $k$ being its modulus.

  The Jacobi elliptic sn function is defined by using the inversion of the first kind  of the incomplete elliptic integral, i.e., the first kind of the incomplete elliptic integral (also called Legendre normal form)~\cite{sn}  \bee\label{inte}
   u=\int_0^{\sigma}\frac{ds}{\sqrt{1-k^2\sin^2s}}, \quad k\in (0, 1) \ene
    generates the Jacobi elliptic sn function
    \bee \label{sn}
    {\rm sn}(u,k)=\sin\sigma, \ene
    from which we have the limit properties: ${\rm sn}(x,k)=\tanh(x)$ for $k\to 1$, and ${\rm sn}(x,k)=\sin(x)$ for $k\to 0$. Moreover we further have another two basic
    Jacobi elliptic cn and dn functions: ${\rm cn}(u,k)=\cos\sigma$, and ${\rm dn}(u,k)=\sqrt{1-k^2\sin^2\sigma}>0$
    from Eq.~(\ref{sn}). These three basic Jacobi elliptic  functions admit these two relations:
    ${\rm sn}^2(u,k)+{\rm cn}^2(u,k)=1$ and $k^2{\rm sn}^2(u,k)+{\rm dn}^2(u,k)=1$. In particular, when $\sigma=\pi/2$ the first kind of the ${\it incomplete}$ elliptic integral (\ref{inte}) becomes the first kind of the ${\it complete}$ elliptic integral
    \bee
         K(k)=\int^{\pi/2}_0\!\!\frac{ds}{\sqrt{1-k^2\sin^2s}}.
    \ene
    These Jacobi elliptic functions have the special values ${\rm sn}(0,k)=0$,\, ${\rm cn}(0,k)={\rm dn}(0,k)=1$,  ${\rm sn}(K,k)=1$,\, ${\rm cn}(K,k)=0,\, {\rm dn}(K,k)=\sqrt{1-k^2}$, and have the periodic properties
    ${\rm sn}(u+4K,k)=-{\rm sn}(u+2K,k)={\rm sn}(u,k)$,\, ${\rm cn}(u+4K,k)=-{\rm cn}(u+2K,k)={\rm cn}(u,k)$,
    ${\rm dn}(u+2K,k)={\rm dn}(u,k)$. Based on the above-mentioned periodic properties of the sn function, we know that $2nK(k)\, (n=0,1,2,...)$ are its zero points~\cite{sn,sn2}, i.e.,
    \bee\label{snzero}
     {\rm sn}[2nK(k),k]\equiv 0, \,\,\,\,  k\in (0, 1),\,\, n=0,1,2,...
     \ene

     Moreover, based on the reciprocals and quotients of three basic functions ${\rm sn}(u,k),\, ${\rm cn}(u,k),\, ${\rm dn}(u,k)$, there are other nine Jacobi elliptic functions: ${\rm ns}(u,k)=1/{\rm sn}(u,k$, ${\rm nc}(u,k)=1/{\rm cn}(u,k)$, ${\rm nd}(u,k)=1/{\rm dn}(u,k)$, ${\rm sc}(u,k)={\rm sn}(u,k)/{\rm cn}(u,k)$, ${\rm sd}(u,k)={\rm sn}(u,k)/{\rm dn}(u,k)$,
     ${\rm cd}(u,k)={\rm cn}(u,k)/{\rm dn}(u,k)$, ${\rm cs}(u,k)={\rm cn}(u,k)/{\rm sn}(u,k)$,
    ${\rm ds}(u,k)={\rm dn}(u,k)/{\rm sn}(u,k)$, ${\rm dc}(u,k)={\rm dn}(u,k)/{\rm cn}(u,k)$~\cite{sn,sn2}.

 It follows from Eq.~(\ref{xx}) that we have $X(x)=\sqrt{\pi}/2[1+{\rm erf}(\sqrt{\omega} x)]\in
(0, \sqrt{\pi})$. The choice of $X(x)$ ensures that
$\Phi_{sn}[X(x), k]\to 0$ for all parameters $\nu,\,k$ and $G$ as $x\to -\infty$. Imposing a zero boundary
condition for $x\to +\infty$ we have the constraint for the parameter $\nu$ in the solution (\ref{SN})
\bee \label{con}
\nu=\frac{2n}{\sqrt{\pi}}K(k),  \ene
based on the property (\ref{snzero}), that is to say, $\Phi_{sn}[X(x), k]\to 0$ for all parameters $k,\,G$, and $\nu$ given by Eq.~(\ref{con}) as $x\to +\infty$.

Therefore, for the given Gaussian nonlinearity and the combination of the harmonic and Gaussian potentials given by Eqs.~(\ref{gg}) and (\ref{vv}),  we obtain a family of exact stationary solutions of Eq.~(\ref{nls})
 \bee\label{solution1}
  \psi_{sn,n}(x,t)\!=\!\frac{2\sqrt{2\sqrt{\omega}}\,nkK(k)}{\sqrt{\pi G}}e^{\frac{\omega x^2}{2}}
  {\rm sn}[\theta_n(x),k]e^{\frac{i\omega t}{2}}, \quad
  \ene
 where $k\in (0,1),\, \theta_n(x)=nK(k)[1+{\rm erf}(\sqrt{\omega} x)],\, n=1,2,3,...$, and $\omega=\gamma^2$ is a positive real number. It can be shown that $|\psi_{sn,n}(x,t)|\to 0$ as $x\to \pm \infty$, which means that the  family of exact stationary solutions of Eq.~(\ref{nls}) is bounded.

\begin{figure}[!ht]
\begin{center}
{\scalebox{0.45}[0.55]{\includegraphics{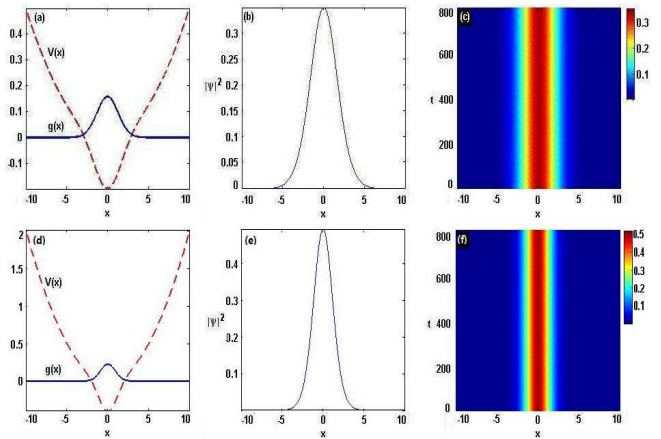}}}
\end{center}
\vspace{-0.2in} \caption{\small (color online)
 (a),(d) The external potential $V(x)$ given by Eq.~(\ref{vv}) (dashed line) and the repulsive interaction $g(x)$ given by Eq.~(\ref{gg}) (solid line) with the Gaussian deeps being $0.16$ for (a) and $0.22$ for (d); (b),(e) The intensity distribution $|\psi_{sn,n}|^2$  of  exact solutions (\ref{solution1}) when $t=0$, in which the widths of the solitons are $12.4$ for (b) and $9$ for (e); (c),(f) The stable analysis of exact solutions (\ref{solution1}) with an initial amplitude stochastic noise of level 0.005. The parameters in the upper (lower) row are $n=1,\,\omega=0.1\, (\omega=0.2),\, k=0.4$, and $G=1$. }
 \label{fig:1}
\end{figure}

 Therefore, we have four free parameters $G,\, n,\, \omega,$ and $k$ for the obtained solutions (\ref{solution1}). In order to investigate the dynamics of exact analytical solutions~(\ref{solution1}),
 we choose the parameters as $G=1,\, n=1,\,\omega=0.1 \, (\omega=0.2)$, and $k=0.4$,  then the profiles of the Gaussian nonlinearity $g(x)$ given by Eq.~(\ref{gg}) (solid line) and the combination of the harmonic and  Gaussian potential $V(x)$ given by Eq.~(\ref{vv}) (dashed line) are plotted in Figs.~\ref{fig:1}a and ~\ref{fig:1}d, among these the nonlinearity can be generated by controlling the Feschbach resonances optically or magnetically using a Gaussian beam~\cite{j5,fr}. In Figs.~\ref{fig:1}b and ~\ref{fig:1}e we show the intensity distribution of the exact solutions (\ref{solution1}) as $t=0$. Moreover, we find that when $\omega$ becomes large ($0.1\to 0.2$), the Gaussian deep for the nonlinearity becomes high ($0.16\to 0.22$), the width of the soliton becomes narrow ($12.4\to 9.0$), and the
 the height of the soliton becomes low ($0.34 \to 0.47$).  Also, we study the stability of the exact solutions in response to perturbation by initial stochastic noise through direct numerical simulations (see Figs.~\ref{fig:1}c and \ref{fig:1}f).

\begin{figure}[!ht]
\begin{center}
{\scalebox{0.43}[0.53]{\includegraphics{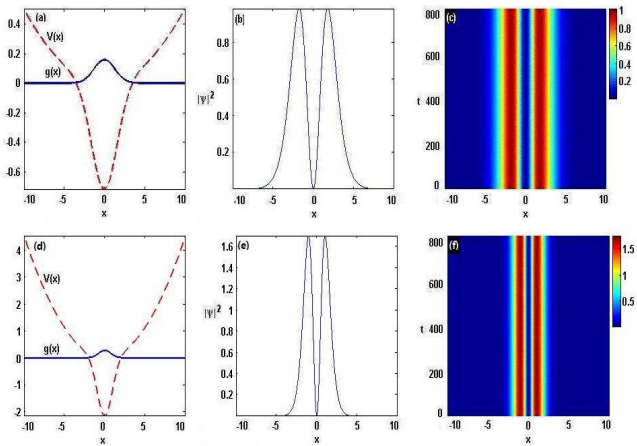}}}
\end{center}
\vspace{-0.2in} \caption{\small (color online)
 (a),(d) The external potential $V(x)$ given by Eq.~(\ref{vv}) (dashed line) and the repulsive interaction $g(x)$ given by Eq.~(\ref{gg}) (solid line) with the Gaussian deeps being $0.16$ for (a) and $0.27$ for (d); (b),(e) the intensity distribution $|\psi_{sn,n}|^2$  of the exact solutions (\ref{solution1}) when $t=0$, in which the widths of the soliton are $14$ for (b) and $8.2$ for (e); (c),(f) the stable analysis of exact solutions (\ref{solution1}) with an initial amplitude  stochastic noise of level 0.005. The parameters in the upper (lower) row are $n=2,\,\omega=0.1\, (\omega=0.3),\, k=0.3$, and $G=1$. }
 \label{fig:2}
\end{figure}

 In the same way, for the case $n=2$, we display two-peak solutions given by Eq.~(\ref{solution1}) in
 Fig.~\ref{fig:2} for other parameters $G=1,\, \omega=0.1\, (\omega=0.3),$ and $k=0.3$. The profiles of the Gaussian nonlinearity $g(x)$ given by Eq.~(\ref{gg}) (solid line) and the combination of the harmonic and  Gaussian potential $V(x)$ given by Eq.~(\ref{vv}) (dashed line) are plotted in Figs.~\ref{fig:2}a and \ref{fig:2}d. Figs.~\ref{fig:2}b and \ref{fig:2}e plot the intensity distribution of the exact solutions (\ref{solution1}) as $t=0$.

 It is observed from Figs.~\ref{fig:1}c, \ref{fig:1}f, \ref{fig:2}c, and \ref{fig:2}f that the exact solutions (\ref{solution1}) with an initial amplitude stochastic noise of level $0.005$ for $n=1$ and $n=2$ are dynamically stable for the chosen parameters. Moreover, for the given parameters, we find that the amplitude noise regimes for the stable solutions are $(0,0.011)$ and $(0,0.03)$ for the solutions (\ref{solution1}) with $n=1$ and $n=2$, respectively. In addition, we also analyze the effect of the phase noise on the solutions such that we find that, for the given parameters, the phase noise regimes for the stable solutions are $(0,0.05)$ and $(0,0.06)$ for the solutions (\ref{solution1}) with $n=1$ and $n=2$, respectively.

 For the given parameters $n,\, G$, and $k$, we analyze the stability of the solutions for the parameter $\omega$, which are related to the external potential, nonlinearity, and solutions. We find that the parameter regimes for the stable solutions are $(0, 0.33)$ and $(0, 0.18)$ for the solutions (\ref{solution1}) with $n=1$ and $n=2$, respectively.

\subsection{The attractive interaction}

For the attractive interaction $G<0$, we have the periodic sd-wave solution of Eq.~(\ref{snls1}) with $X(x,t)\to X(x)$~\cite{yan06, yan}
\bee \label{SD}
\Phi_{sd}[X(x)]=\nu k\sqrt{\frac{2(k^2-1)}{G}}\,{\rm sd}[\nu X(x),k],
\ene
where $k\in (0,1)$, $\nu$ is a constant, the function sd denotes the Jacobi elliptic function, i.e., ${\rm sd}[\nu X(x),k]={\rm sn}[\nu X(x),k]/{\rm dn}[\nu X(x),k]$ (c.f. Eq.~(\ref{sn}))~\cite{sn}, and the condition is required
\bee \label{epsilon}
\varepsilon=\nu^2(1-2k^2).
\ene

Since ${\rm dn}[\nu X(x), k]>0$ for all $X$ and $k$, thus we know that ${\rm sd}(\nu X(x), k)={\rm sn}(\nu X(x), k)/{\rm dn}(\nu X(x), k)$ has the same zero points as ${\rm sn}[\nu X(x), k]$, i.e., ${\rm sd}[2nK(k), k]=0\, (n=0,1,2,...)$. Similarly, based on the constraint for $\nu$ given by Eq.~(\ref{con}) for the requirement of the zero boundary condition for $x\to \pm\infty$, for the combination of harmonic and Gaussian potentials given by Eq.~(\ref{vv}), we have another family of exact solutions of Eq.~(\ref{nls})
 \bee \label{solution2}
\nonumber \psi_{sd,n}(x,t)= \frac{2nkK(k)\sqrt{2\sqrt{\omega}(1-k^2)}}{\sqrt{-\pi G}}\,e^{\frac{\omega x^2}{2}} \\
\times {\rm sd}[\theta_n(x),k]e^{\frac{i\omega t }{2}},\qquad
\ene
where $k\in (0,1),\, \theta_n(x)=nK(k)[1+{\rm erf}(\sqrt{\omega} x)],\,  n=1,2,3,...$, and $\omega$ is a positive real number. It can be shown that $|\psi_{sd,n}(x,t)|\to 0$ as $x\to \pm \infty$, which means that the  family of exact stationary solutions of Eq.~(\ref{nls}) is bounded.

The required condition (\ref{epsilon}) leads to the following three cases for different modulus $k\in (0,1)$:
\bee\label{ecase} \begin{array}{rlll}
{\rm I}: & \quad \varepsilon>0 &\quad {\rm for}& \quad 0<k<1/\sqrt{2}, \vspace{0.05in}\cr
{\rm II}: &\quad \varepsilon=0 &\quad {\rm for} &\quad k=1/\sqrt{2},   \vspace{0.05in}\cr
{\rm III}: &\quad \varepsilon<0 &\quad {\rm for} &\quad 1/\sqrt{2}<k<1,  \end{array} \ene
which make the external potential $V(x)$ given by Eq.~(\ref{vv}) admit three different types of potentials, i.e., the harmonic-like potential ($\varepsilon>0$, see Figs.~\ref{fig:3} and \ref{fig:4}), the harmonic potential ($\varepsilon=0$), and the double-well potential ($\varepsilon<0$, see Figs.~\ref{fig:3a} and \ref{fig:4a}). Exact solutions of Eq.~(\ref{nls}) with the harmonic potential case ($\varepsilon=0$) have been studied~\cite{bb07}. Here we consider solutions of Eq.~(\ref{nls}) with other two cases ($\varepsilon>0$ and $\varepsilon<0$).

\subsubsection{The case  $\varepsilon>0$}

\begin{figure}
\begin{center}
{\scalebox{0.45}[0.55]{\includegraphics{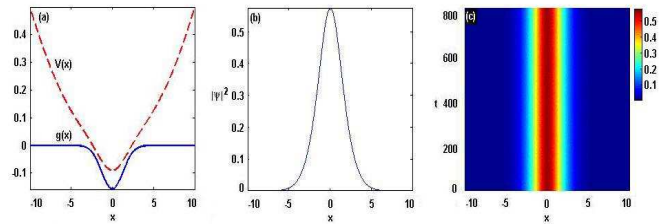}}}
\end{center}
\vspace{-0.2in} \caption{\small (color online)
(a) The external potential $V(x)$ given by Eq.~(\ref{vv}) (dashed line) and the attractive interaction $g(x)$ given by Eq.~(\ref{gg}) (solid line) with the Gaussian deep being $0.15$; (b) the intensity distribution $|\psi_{sd,n}|^2$  of the exact solutions (\ref{solution2}) when $t=0$, in which the width of the soliton is $12$; (c) the stable analysis of the exact solutions (\ref{solution2}) with an initial amplitude  stochastic noise of level 0.005. The parameters are $n=1,\,\omega=0.1,\,k=0.5$, and $G=-1$. }
 \label{fig:3}
\end{figure}

 In order to investigate the dynamics of exact analytical solutions~(\ref{solution2}),
 we choose the parameters as $G=-1,\, n=1,\,\omega=0.1$, and $k=0.5$ satisfying  Case I given by Eq.~(\ref{ecase}), i.e., $0<k<1/\sqrt{2}$. Thus the profiles of the Gaussian nonlinearity $g(x)$ given by Eq.~(\ref{gg}) (solid line) and the combination of harmonic and Gaussian potentials $V(x)$ given by Eq.~(\ref{vv}) (dashed line) are plotted in Fig.~\ref{fig:3}a, the intensity distribution of the exact solutions (\ref{solution2}) for $t=0$ is displayed in Fig.~\ref{fig:3}b, and the numerical result for the evolution of the exact solutions (\ref{solution2}) is shown in Fig.~\ref{fig:3}c.

\begin{figure}
\begin{center}
{\scalebox{0.45}[0.55]{\includegraphics{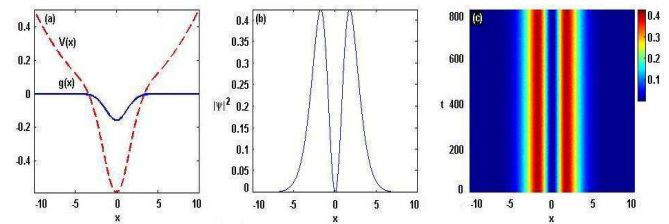}}}
\end{center}
\vspace{-0.2in} \caption{\small (color online)
(a) The external potential $V(x)$ given by Eq.~(\ref{vv}) (dashed line) and the attractive interaction $g(x)$ given by Eq.~(\ref{gg}) (solid line) with the Gaussian deep being $0.15$; for the parameters $n=2,\,\omega=0.1,\, k=0.2$, and $G=-1$, (b) the intensity  distribution $|\psi_{sd,n}|^2$  of the exact solutions (\ref{solution2}) when $t=0$, in which the width of the soliton is $13.6$; (c) the stable analysis of the exact solutions (\ref{solution2}) with an initial amplitude  stochastic noise of level 0.005.}
 \label{fig:4}
\end{figure}

 Similarly, for the case $n=2$, we display the two-peak solutions given by Eq.~(\ref{solution2}) in Fig.~\ref{fig:4} for these parameters $\omega=0.1,\, k=0.2$, and $G=-1$.

 It is observed from Figs.~\ref{fig:3}c and \ref{fig:4}c that the exact solutions (\ref{solution2}) for $n=1$ and $n=2$ are dynamically stable for the chosen parameters. Moreover, for the given parameters, we find that the amplitude noise regimes for the stable solutions are $(0, 0.02)$ and $(0,0.018)$ for the solutions (\ref{solution2}) with $n=1$ and $n=2$, respectively. In addition, we also analyze the effect of the phase noise on the solutions such that we find that, for the given parameters, the phase noise regimes for the stable solutions are $(0, 0.07)$ and $(0,0.05)$ for the solutions (\ref{solution2}) with $n=1$ and $n=2$, respectively.

 For the given parameters $n,\, G$, and $k$, we analyze the stability of the solutions for the parameter $\omega$, which are related to the external potential, nonlinearity, and solutions. We find that the parameter regimes for the stable solutions are $(0, 0.2)$ and $(0, 0.46)$ for the solutions (\ref{solution2}) with $n=1$ and $n=2$, respectively.

\subsubsection{The case  $\varepsilon<0$}

\begin{figure}
\begin{center}
{\scalebox{0.45}[0.56]{\includegraphics{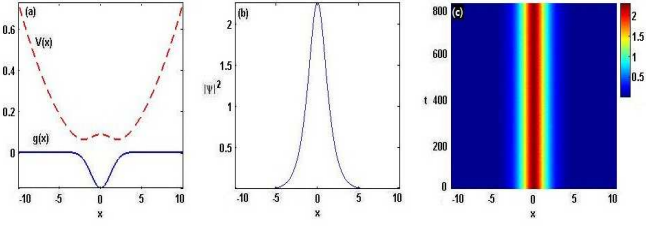}}}
\end{center}
\vspace{-0.2in} \caption{\small (color online)
(a) The double-well potential $V(x)$ given by Eq.~(\ref{vv}) (dashed line) and the attractive interaction $g(x)$ given by Eq.~(\ref{gg}) (solid line) with the Gaussian deep being $0.17$; (b) the intensity distribution $|\psi_{sd,n}|^2$  of the exact solutions (\ref{solution2}) when $t=0$,  in which the width of the soliton being $11$; (c) the stable analysis of the exact solutions (\ref{solution2}) with an initial stochastic noise of level 0.005. The parameters are $n=1,\,\omega=0.12,\,k=0.8$, and $G=-1$.}
 \label{fig:3a}
\end{figure}

 In this case, we choose the parameters as $G=-1,\, n=1,\,\omega=0.12$, and $k=0.8$ satisfying Case III given by Eq.~(\ref{ecase}), i.e., $1/\sqrt{2}<k<1$, to investigate the dynamics of exact analytical solutions~(\ref{solution2}). Thus the profiles of the Gaussian nonlinearity $g(x)$ given by Eq.~(\ref{gg}) (solid line) and the double-well potential $V(x)$ given by Eq.~(\ref{vv}) (dashed line) are plotted in Fig.~\ref{fig:3a}a, the intensity distribution of the exact solutions (\ref{solution2}) for $t=0$ is displayed in Fig.~\ref{fig:3a}b, and the numerical result for the evolution of the exact solutions (\ref{solution2}) is shown in Fig.~\ref{fig:3a}c.

\begin{figure}
\begin{center}
{\scalebox{0.45}[0.56]{\includegraphics{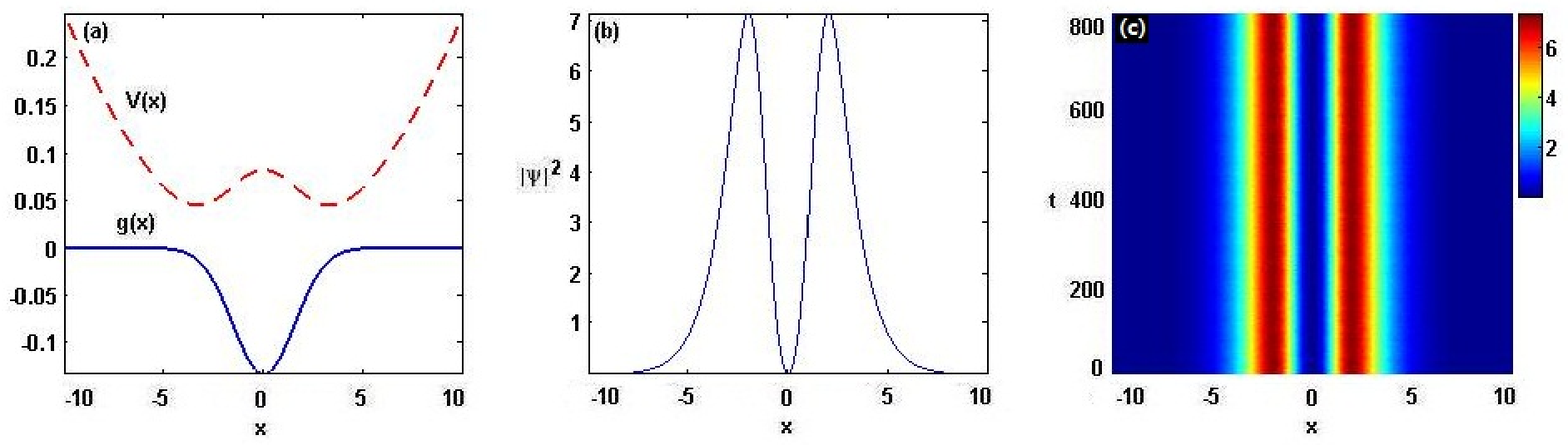}}}
\end{center}
\vspace{-0.2in} \caption{\small (color online)
(a) The double-wel potential $V(x)$ given by Eq.~(\ref{vv})(dashed line) and the attractive interaction $g(x)$ given by Eq.~(\ref{gg}) (solid line) with the Gaussian deep being $0.13$; (b) the intensity  distribution $|\psi_{sd,n}|^2$  of the exact solutions (\ref{solution2}) when $t=0$,  in which the width of the soliton being $16$; (c) the stable analysis of the exact solutions (\ref{solution2}) with an initial stochastic noise of level 0.005. The parameters are $n=2,\,\omega=0.07,\, k=0.75$, and $G=-1$.}
 \label{fig:4a}
\end{figure}

Similarly, for the case $n=2$, we display the two-peak solutions given by Eq.~(\ref{solution2}) in Fig.~\ref{fig:4a} for these parameters $\omega=0.07,\, k=0.75$, and $G=-1$ also satisfying Case III given by Eq.~(\ref{ecase}), i.e., $1/\sqrt{2}<k<1$.

 It is observed from Figs.~\ref{fig:3a}c and \ref{fig:4a}c that the exact solutions (\ref{solution2}) for $n=1$ and $n=2$ are dynamically stable for the chosen parameters. Moreover, for the given parameters, we find that the amplitude noise regimes for the stable solutions are $(0, 0.04)$ and $(0,0.05)$ for the solutions (\ref{solution2}) with $n=1$ and $n=2$, respectively. In addition, we also analyze the effect of the phase noise on the solutions such that we find that, for the given parameters, the phase noise regimes for the stable solutions are $(0, 0.05)$ and $(0,0.04)$ for the solutions (\ref{solution2}) with $n=1$ and $n=2$, respectively.

 For the given parameters $n,\, G$, and $k$, we analyze the stability of the solutions for the parameter $\omega$, which are related to the external potential, nonlinearity, and solutions. We find that the parameter regimes for the stable solutions are $(0, 0.13)$ and $(0, 0.07)$ for the solutions (\ref{solution2}) with $n=1$ and $n=2$, respectively.

\section{Time-depended potential and nonlinearity}

   So far we have considered stationary solutions of the GP equation with the potential and nonlinearity depending only on the spatial coordinate. Now we focus on the case that the potential $V(x,t)$ and nonlinearity $g(x,t)$ depend on both time and spatial coordinates, and the gain or loss term $\Gamma(t)$ is a function of time (see Eq.~(\ref{nls1})).

\begin{figure}
\begin{center}
\vspace{0.02in}
{\scalebox{0.43}[0.48]{\includegraphics{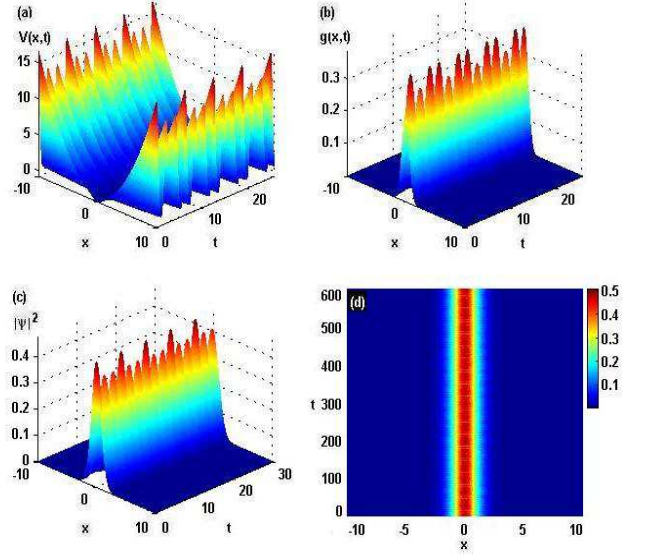}}}
\end{center}
\vspace{-0.2in} \caption{\small (color online)
(a) The external potential $V(x,t)$ given by Eq.~(\ref{v});  (b) the repulsive interaction $g(x,t)$ given by Eq.~(\ref{g}) with the maximal Gaussian deep being $0.38$; (c) the intensity distribution $|\psi_{sn,n}|^2$  of the exact solutions (\ref{solution3}) with the width of the soliton being $6.2$; (d) the evolution of the exact solutions (\ref{solution3}) with an initial amplitude stochastic noise of level $0.005$. The parameters are $n=1,\,\gamma(t)=6[0.1+0.01\cos(t)],\,\Gamma(t)=-0.3\sin(3t),\,k=0.3$, and $G=1$.}
 \label{fig:5}
\end{figure}

\begin{figure}
\begin{center}
\vspace{0.02in}
{\scalebox{0.43}[0.48]{\includegraphics{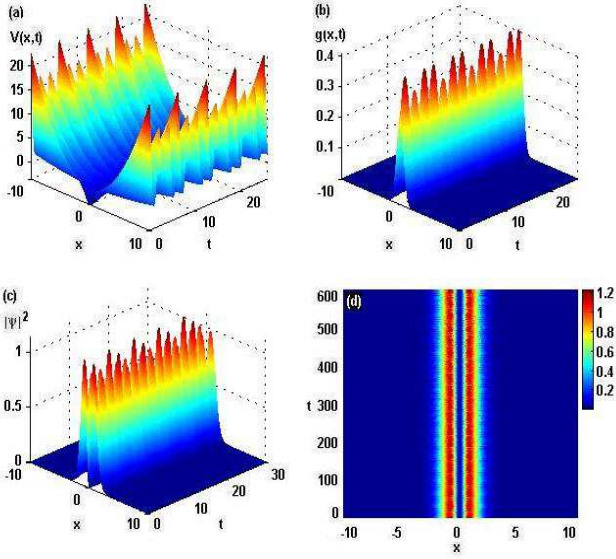}}}
\end{center}
\vspace{-0.2in} \caption{\small (color online)
(a) The external potential $V(x,t)$ given by Eq.~(\ref{v});  (b) the repulsive interaction $g(x,t)$ given by Eq.~(\ref{g}) with the maximal Gaussian deep being $0.41$; (c) the intensity  distribution $|\psi_{sn,n}|^2$  of the exact solutions (\ref{solution3}) with the width of the soliton being $6.8$; (d) the evolution of the exact solutions (\ref{solution3}) with an initial amplitude stochastic noise of level $0.005$. The parameters  are $n=2,\, \gamma(t)=6.4[0.1+0.01\cos(t)],\,\Gamma(t)=-0.3\sin(3t),\,k=0.2$, and $G=1$.}
 \label{fig:6}
\end{figure}

 \subsection{The repulsive interaction}

 For the determined potential $V(x,t)$ and nonlinearity $g(x,t)$ and other conditions given by Eqs.~(\ref{v}) and (\ref{sys}), we have a family of exact solutions of Eq.~(\ref{nls1}) with the repulsive interaction $G>0$ by using the solution of Eq.~(\ref{snls1}) given by Eq.~(\ref{SN}) and the determined similarity transformation (\ref{trans})
\bee\label{solution3}
  \psi_{sn,n}(x,t)=\frac{2\sqrt{2\gamma(t)}\,nkK(k)}{\sqrt{\pi G}}\,e^{\frac{\gamma^2(t)x^2}{2}+\int \Gamma(t)dt}\nonumber\\
  \times {\rm sn}[\theta_n(x,t),k]e^{i\varphi (x,t)},
  \ene
 where the function sn denotes the Jacobi elliptic function with $k\in (0,1)$ being its modulus, \,$K(k)=\int^{\pi/2}_0[1-k^2\sin^2(s)]^{-1/2}ds$, $\theta_n(x,t)=nK(k)[1+{\rm erf}(\gamma(t) x)],\, n=1,2,3,...$, and the phase $\varphi(x,t)$ is given by Eq.~(\ref{q1}).

To make sure that the nonlinearity $g(x,t)$ and the coefficients of the external potential $V(x,t)$ in space
are bounded for realistic cases, we choose the inverse of the width of the localized solution $\gamma(t)$ and the gain or loss distribution $\Gamma(t)$ as the localized periodic functions in the form
\bee\label{tpara}
\begin{array}{l}
\gamma(t)=6[0.1+0.01\cos(t)],\vspace{0.1in}\cr
\Gamma(t)=-0.3\sin(3t),\vspace{0.1in}\cr
k=0.3, \quad G=1.
\end{array}
\ene
For the case $n=1$, we display the one-peak solutions given by Eq.~(\ref{solution3}).
In Fig.~\ref{fig:5}, we plot the potential $V(x,t)$ given by Eq.~(\ref{v}), the nonlinearity $g(x,t)$ given by Eq.~(\ref{g}), the intensity distribution of exact solutions (\ref{solution3}), and the numerical result for the evolution of exact solutions (\ref{solution3}) for the chosen parameters.

 Similarly, for the case $n=2$, we display the two-peak solutions given by Eq.~(\ref{solution3}) in
Fig.~\ref{fig:6} for other chosen parameters $\gamma(t)=6.4[0.1+0.01\cos(t)],\,\Gamma(t)=-0.3\sin(3t),\,k=0.2$, and $G=1$.

It is observed that the amplitude of the potential $V(x,t)$ varies
quasi-periodically with respect to time, so does the nonlinearity $g(x,t)$. It is also observed from Figs.~\ref{fig:5}d and \ref{fig:6}d that exact solutions (\ref{solution3}) for $n=1$ and $n=2$ are dynamically stable for the chosen parameters. Moreover, for the given parameters, we find that the amplitude noise regimes for the stable solutions are $(0, 0.011)$ and $(0,0.015)$ for the solutions (\ref{solution3}) with $n=1$ and $n=2$, respectively. In addition, we also analyze the effect of the phase noise on the solutions such that we find that, for the given parameters, the phase noise regimes for the stable solutions are $(0, 0.05)$ and $(0,0.04)$ for the solutions (\ref{solution3}) with $n=1$ and $n=2$, respectively.

For the given parameters $n,\, G,\, k$, and $\gamma(t)$ given by Eq.~(\ref{tpara}), we analyze the stability of the solutions for the varying gain or loss term
\bee \label{gain}
 \Gamma(t)=c\sin(3t) \qquad (c={\rm const}),
  \ene
  which are related to the external potential, nonlinearity, and solutions.
We find that the parameter ($c)$ regimes for the stable solutions are $(-0.3, 0)$ and $(-0.3, 0)$ for the solutions (\ref{solution3}) with $n=1$ and $n=2$, respectively.

\begin{figure}
\begin{center}
\vspace{0.05in}
{\scalebox{0.43}[0.48]{\includegraphics{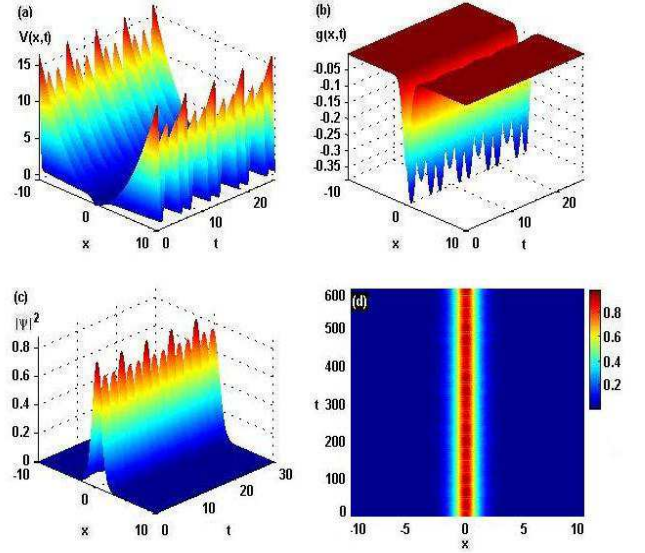}}}
\end{center}
\vspace{-0.2in} \caption{\small (color online)
(a) The potential $V(x,t)$ given by Eq.~(\ref{v}); (b) the attractive interaction $g(x,t)$ given by Eq.~(\ref{g}) with the maximal Gaussian deep being $0.38$; (c) the intensity distribution $|\psi_{sd,n}|^2$ of exact solutions (\ref{solution4}) with the width of the soliton being $5.8$; (d) the evolution of the exact solutions (\ref{solution4}) with an initial amplitude stochastic noise of level 0.005. The parameters are $n=1,\,\gamma(t)=6[0.1+0.01\cos(t)],\,\Gamma(t)=-0.3\sin(3t),\,k=0.4$, and $G=-1$.}
 \label{fig:7}
\end{figure}

\begin{figure}
\begin{center}
\vspace{0.05in}
{\scalebox{0.43}[0.48]{\includegraphics{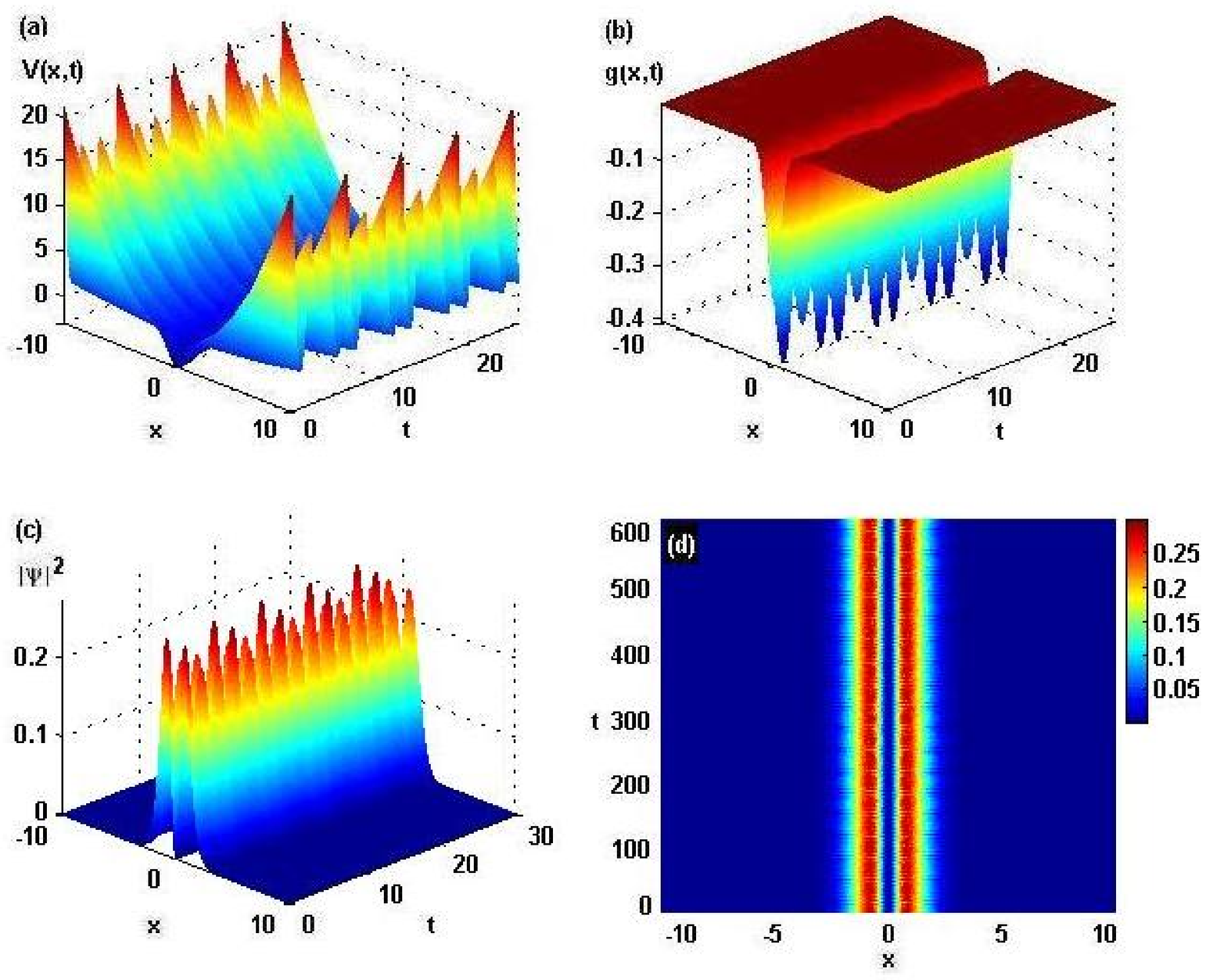}}}
\end{center}
\vspace{-0.2in} \caption{\small (color online)
(a) The potential $V(x,t)$ given by Eq.~(\ref{v}); (b) the attractive interaction $g(x,t)$ given by Eq.~(\ref{g}) with the maximal Gaussian deep being $0.4$, (c) the intensity distribution $|\psi_{sd,n}|^2$ of exact solutions (\ref{solution4}) with the width of the soliton being $6.4$; (d) the evolution of the exact solutions (\ref{solution4}) with an initial amplitude stochastic noise of level 0.005. The parameters are $n=2,\, \gamma(t)=6.3[0.1+0.01\cos(t)],\, \Gamma(t)=-0.3\sin(3t),\, k=0.1$, and $G=-1$.}
 \label{fig:8}
\end{figure}

\subsection{The attractive interaction}

 We have a family of exact solutions of Eq.~(\ref{nls1}) with the
attractive interaction $G<0$ by using the solution (\ref{SD}) of Eq.~(\ref{snls1}) and the similarity transformation (\ref{trans})
\bee\label{solution4}
  \psi_{sd,n}(x,t)=\frac{2\sqrt{2\gamma(t)(1-k^2)}\,nkK(k)}{\sqrt{-\pi G}}\,e^{\frac{\gamma^2(t) x^2}{2}}\nonumber\\
  \times e^{\int \Gamma(t)dt}\,{\rm sd}[\theta_n(x,t),k]e^{i\varphi (x,t)},
  \ene
 where the sd function ${\rm sd}(\cdot, k)\equiv {\rm sn}(\cdot, k)/{\rm dn}(\cdot, k)$ denotes the Jacobi elliptic function with $k\in (0,1)$ being its modulus,\,$K(k)=\int^{\pi/2}_0[1-k^2\sin^2s]^{-1/2}ds$, $\theta_n(x,t)=nK(k)[1+{\rm erf}(\gamma(t) x)],\, n=1,2,3,...$, and the phase $\varphi(x,t)$ is defined by Eq.~(\ref{q1}).

In Fig.~\ref{fig:7}, we plot the potential $V(x,t)$ given by Eq.~(\ref{v}), the attractive interaction $g(x,t)$ given by Eq.~(\ref{g}), the intensity distribution of the exact solutions (\ref{solution4}) and the numerical result for the evolution of the exact solutions (\ref{solution4}) for the chosen parameters
\bee \label{tpara2}
\begin{array}{l}
n=1,\,\, \,\, G=-1,\,\,\,\, k=0.4, \vspace{0.1in} \cr
\gamma(t)=6[0.1+0.01\cos(t)],\vspace{0.1in}\cr
\Gamma(t)=-0.3\sin(3t).
\end{array}
\ene

Similarly, for the case $n=2$, we show the two-peak solutions given by Eq.~(\ref{solution4}) in Fig.~\ref{fig:8} for another parameters $\gamma(t)=6.3[0.1+0.01\cos(t)],\,\Gamma(t)=-0.3\sin(3t),\, k=0.1$, and $G=-1$.

 It is  observed that the amplitude of the potential $V(x,t)$ varies quasi-periodically with respect to time, so does the nonlinearity $g(x,t)$. It is also observed from Figs.~\ref{fig:7}d and \ref{fig:8}d that the exact solutions (\ref{solution4}) for $n=1$ and $n=2$ are dynamically stable for the chosen parameters. Moreover, for the given parameters, we find that the amplitude noise regimes for the stable solutions are $(0, 0.015)$ and $(0,0.01)$ for the solutions (\ref{solution4}) with $n=1$ and $n=2$, respectively. In addition, we also analyze the effect of the phase noise on the solutions such that we find that, for the given parameters, the phase noise regimes for the stable solutions are $(0, 0.05)$ and $(0,0.04)$ for the solutions (\ref{solution4}) with $n=1$ and $n=2$, respectively.

For the given parameters $n,\, G,\, k$, and $\gamma(t)$ given by Eq.~(\ref{tpara2}), we analyze the stability of the solutions for the gain or loss term given by Eq.~(\ref{gain}), which are related to the external potential, nonlinearity, and solutions.
We find that the parameter ($c)$ regimes for the stable solutions are $(-0.3, 0)$ and $(-0.3, 0)$ for the solutions (\ref{solution4}) with $n=1$ and $n=2$, respectively.

\section{Conclusions}

In conclusion,  we have studied matter-wave solutions in quasi-one-dimensional Bose-Einstein condensates in the combination of the harmonic and Gaussian potentials. As a consequence, we have found several families of exact solutions of the GP equation with the combination of the harmonic and Gaussian potentials by using the similarity transformation.
Moreover, the stability of the obtained exact solutions is investigated by using numerical simulations such that some stable solutions are found. In particular, we analyze the effects of both the amplitude noise and the phase noise. We also give the parameter regimes for the stable solutions. These results may raise the possibility of relative experiments and potential applications.

The used method can be also extended to investigate exact solutions of some other physical models in BECs such as the cubic-quintic, the two-component, the spinor-1, and/or higher-dimensional GP equations with the combination of the harmonic and Gaussian potentials. In fact, our idea can also be applied to the related NLS equation with varying coefficients and its extensions in nonlinear optics~\cite{b1,b4,j8}.

\acknowledgments

This work was supported by the NSFC under Grant Nos. 11071242 and 61178091.


\begin{thebibliography}{99} \small

\bibitem{b1} C. Sulem and P. L. Sulem, {\em The Nonlinear Schr\"oinger Equation:
Self-focusing and Wave Collapse } (Springer-Verlag, New York, 1999).

\bibitem{b2} M. J. Ablowitz and P. A. Clarkson {\em Solitons, Nonlinear Evolution Equations and Inverse scattering}
(Combridge University Press, Cambridge, 1991).

\bibitem{b3}  B. A. Malomed, {\em  Solition Management in Periodic Systems} (Sringer, New York, 2006).

\bibitem{b4} B. A. Malomed, D. Mihalache, F. Wise, and L. Torner, J. Opt.
B: Quantum Semiclassical Opt. {\bf 7}, R53 (2005).

\bibitem{b5} R. Carretero-Gonz\'alez, D. J. Frantzeskakis, and P. G. Kevrekidis,
Nonlinearity {\bf 21}, R139 (2008); D. J. Frantzeskakis, J. Phys. A {\bf 43}, 213001 (2010).

\bibitem{b6} V. G. Ivancevic, Cognitive Computation  {\bf 2}, 17 (2010);
Z. Y. Yan, Phys. Lett. A {\bf 375}, 4274 (2011); {\it ibid}, Commun. Thero. Phys. {\bf 54}, 947 (2010).

\bibitem{bb01}H.-H. Chen and C.-S. Liu, Phys. Rev. Lett. {\bf 37}, 693 (1976).

\bibitem{bb02} V. V. Konotop, Phys. Rev. E {\bf 47}, 1423 (1993); V. V.
Konotop, O. A. Chubykalo, and L. Vazquez, {\it ibid}. {\bf 48}, 563 (1993).

\bibitem{bb03} V. N. Serkin and A. Hasegawa, Phys. Rev. Lett. {\bf 85}, 4502
(2000); V. N. Serkin and T. L. Belyaeva, {\it ibid}. {\bf 74}, 573 (2001); V. I. Kruglov, A. C. Peacock, and J. D. Harvey, {\it ibid}. {\bf 90}, 113902 (2003); S. A. Ponomarenko and G. P. Agrawal, Phys. Rev. Lett. {\bf 97}, 013901 (2006); V. N. Serkin, A. Hasegawa, and T. L. Belyaeva, Phys. Rev. Lett. {\bf 98}, 074102 (2007); A. Kundu, Phys. Rev. E {\bf 79}, 015601R (2009).

\bibitem{j8} Y. Kivshar and G. P. Agrawal, {\em Optical Solitons: From
Fibers to Photonic Crystals} (Academic Press, New
York, 2003).

\bibitem{bb04} V. N. Serkin and A. Hasegawa, IEEE J. Sel. Top. Quantum
Electron. {\bf 8}, 418 (2002); V. N. Serkin, A. Hasegawa, and
T. L. Belyaeva, Phys. Rev. Lett. {\bf 92}, 199401 (2004).

\bibitem{j5} L. P. Pitaevskii and S. Stringari, {\em Bose-Einstein
Condensation} (Oxford University Press, Oxford, 2003); F. Dalfovo,
S. Giorgini, L. P. Pitaevskii, and S. Stringari, Rev. Mod.
Phys. {\bf 71}, 463 (1999); A. J. Leggett, Rev. Mod. Phys. {\bf 73}, 307 (2001);
E. Timmermans {\it et al.,} Phys. Rep. {\bf 315}, 199 (1999). O. Morsch and M. Oberthaler, Rev. Mod. Phys. {\bf 78}, 179 (2006).

\bibitem{j55} F. Kh. Abdullaev {\it et al.,} Phys. Rev. Lett. {\bf 90}, 230402 (2003);
  V. A. Brazhnyi and V. V. Konotop, Mod. Phys. Lett. B {\bf 18}, 627 (2004).

\bibitem{j6} P. G. Kevrekidis, D. J. Frantzeskakis, and R. Carretero-Gonzalez, {\em  Emergent Nonlinear Phenomena in Bose-Einstein Condensates: Theory and Experiment} (Springer, New York, 2008).

\bibitem{fr} A. J. Moerdijk, B. J. Verhaar, and A. Axelsson, Phys. Rev. A {\bf 51}, 4852 (1995);
S. Inouye {\it et al.,}  Nature (London) {\bf 392}, 151
(1998); C. Chin {\it et al.,} Rev. Mod. Phys. {\bf 82}, 1225 (2010).


\bibitem{j7} Y. V. Kartashov, B. A. Malomed, and L. Torner, Rev. Mod. Phys. {\bf 83}, 247 (2011).


\bibitem{liang} Z. X. Liang, Z. D. Zhang, and W. M. Liu, Phys. Rev. Lett. {\bf 94},
050402 (2005).

\bibitem{bb07} J. Belmonte-Beitia, V. M. P\'erez-Garc\'ia, V. Vekslerchik, and P. J.
 Torres, Phys. Rev. Lett. {\bf 98}, 064102 (2007).

\bibitem{bb08} J. Belmonte-Beitia, V. M. P\'erez-Garc\'ia, V. Vekslerchik, and V.
V. Konotop, Phys. Rev. Lett. {\bf 100}, 164102 (2008).

\bibitem{yan11} Z. Y. Yan, X. F. Zhang, and W. M. Liu, Phys. Rev. A {\bf 84}, 023627 (2011).

\bibitem{yu9} H. Friedrich, G. Jacoby, and C. G. Meister, Phys. Rev. A {\bf 65}, 032902 (2002);
Yu. V. Bludov,  Z. Y. Yan,  V. V. Konotop, Phys. Rev. A {\bf 81}, 063610 (2010).

\bibitem{yan06} Z. Y. Yan and V. V. Konotop, Phys. Rev. E {\bf 80}, 036607 (2009); Z. Y. Yan and C. Hang, Phys. Rev. A {\bf 80}, 063626 (2009).

\bibitem{vvk} V. M. P\'erez-Garc\'ia,  P. J. Torres, and V.
V. Konotop, Physica D {\bf 221}, 31 (2006).

\bibitem{yan12} Z. Y. Yan, V. V. Konotop, A. V. Yulin, and W. M. Liu, Phys. Rev. E {\bf 85}, 016601 (2012).

\bibitem{low} A. I. Safonov {\it et al.}, Phys. Rev. Lett. {\bf 81}, 4545 (1998);  B. P. Anderson and M. Kasevich, Science {\bf 282}, 1686 (1998); A. G\"rlitz {\it et al.}, Phys. Rev. Lett. {\bf 87}, 130402 (2001);
     C. Menotti and S. Stringari, Phys. Rev. A {\bf 66}, 043610 (2002);
    N. G. Parker, N. P. Proukakis, M. Leadbeater, and C. S. Adams, Phys. Rev. Lett. {\bf 90}, 220401 (2003);
   F. Kh. Abdullaev and J. Garnier, Phys. Rev. A {\bf 70}, 053604 (2004); S. Wildermuth {\it et al.,} Nature {\bf  435}, 440 (2005); J. Billy {\it et al.,} Nature {\bf 453}, 891 (2008).


\bibitem{gl} B. Kneer {\it et al.,} Phys. Rev. A {\bf 58}, 4841 (1998);
 P. D. Drummond and K. V. Kheruntsyan, Phys. Rev. A {\bf 63}, 013605 (2000);
M. K\"ohl {\it et al.,} Phys. Rev. Lett. {\bf 88}, 080402 (2002);
V. I. Kruglov, M. K. Olsen, and M. J. Collett, Phys. Rev. A {\bf 72}, 033604 (2005);
A. P. D. Love {\it et al.,} Phys. Rev. Lett. {\bf 101}, 067404 (2008);
N. Syassen, {\it et al.,}  Science {\bf 320}, 1329 (2008);
S. Rajendran, M. Lakshmanan, and P. Muruganandam, J. Math. Phys. {\bf 52}, 023515 (2011).

\bibitem{lan} G. Whitham, {\em Linear and Nonlinear Waves} (Wiley-Interscience, New York, 1974).

\bibitem{exp00} K. E. Strecker {\it et al.,}  Nature {\bf 417}, 150 (2002).

\bibitem{exp0} T. Schumm {\it et al.,} Nature Phys. {\bf 1}, 57 (2005).

\bibitem{exp1} C. Becker {\it et al.,} Nature Phys.  {\bf 4}, 496 (2008).

\bibitem{exp2} A. Weller {\it et al.,} Phys. Rev. Lett. {\bf 101}, 130401 (2008); S. Stellmer {\it et al.,} Phys. Rev. Lett. {\bf 101}, 120406 (2008);

\bibitem{yan} Z. Y. Yan, Phys. Lett. A {\bf 374}, 4838 (2010); {\it ibid}, {\bf 333}, 193 (2004);
Phys. Scr. {\bf 78}, 035001 (2008).

\bibitem{gau}  Y. Shin {\it et al.,}  Phys. Rev. Lett. {\bf 92}, 150401 (2004).

\bibitem{pu} S. K. Adhikari, H. Lu, and H. Pu, Phys. Rev. A {\bf 80}, 063607 (2009).

\bibitem{sn} M. Abramowitz and I. A. Stegun, {\em Handbook of Mathematical Functions} (Dover Publications, Inc., New York, 1965).

\bibitem{sn2} E. T. Whittaker and G. N. Watson, {\em A Course of Modern Analysis} (Cambridge University Press, New York, 4th edition, 1962).

 \end{thebibliography}
  \end{document}